\documentclass[twocolumn,showpacs,showkeys,preprintnumbers,amsmath,amssymb]{revtex4}

\usepackage{graphicx}

\preprint{}

\begin{document}

\title{$\phi$ meson production in near threshold proton-nucleus collisions}

\author{H.W. Barz$^1$ and M. Z\'et\'enyi$^{2,1}$}
\affiliation{
$^1$ Forschungszentrum Rossendorf, Pf 510119, D-01314 Dresden, Germany\\
$^2$ KFKI Research Institute for Particle and Nuclear Physics, POB 49,
H-1525 Budapest, Hungary
}

\date{}

\begin{abstract}
  The cross section for production of $\phi$ mesons in 
  proton-nucleus reactions is calculated 
  as a function of the target mass. The decay
  width of the $\phi$ meson is affected by the change of the masses of
  the $\phi$, $K^+$ and $K^-$ mesons in the medium. A strong
  attractive $K^-$ potential leads to a measurable change
  of the behavior of the cross section as a function of of the 
  target mass. Comparison between the kaon and electron decay modes  
  are made.
\end{abstract}


\maketitle

\section{Introduction}

Particle production in near threshold $p+A$ reactions is a sensitive
probe of the in-medium modification of the particle properties. A
change of the particle spectral function which is mainly characterized
by the particle mass may result in a drastic change of the
available phase space and hence the production cross section. Decay
widths of the produced particles may be influenced in a similar way.

Kaonic atom experiments \cite{Katom} and the enhanced $K^-$ production
observed in 
heavy-ion collisions by the FOPI \cite{fopi1,fopi2} and the KaoS
\cite{kaos1,kaos3} collaborations point to a strong
attractive $K^-$ potential. These results seemed to support
the predictions of early theoretical approaches based on effective
chiral Lagrangians \cite{Nelson-Kaplan}. However, more sophisticated
theoretical calculations \cite{Kolomeitsev,SibCas,Schaffner,Cieply,Lutz}
lead to weaker or strongly momentum dependent $K^-$ potentials, 
which can even become repulsive for large momenta and densities. 
On the contrary $K^{+}$ and $\phi$ potentials (repulsive for $K^{+}$
and attractive for the $\phi$ meson) are believed to 
depend only weakly on the density of nuclear matter.

The study of $\phi$ meson production in $p+A$ collisions provides an
independent test of the in-medium kaon and $\phi$ potentials. In case
of a strong attractive $K^-$ potential and moderate $K^+$ and $\phi$
potentials the mean proper lifetime of $\phi$ mesons decreases
from the vacuum value of about 50 fm/$c$ to an order of
magnitude smaller value in normal nuclear matter. Therefore, $\phi$
mesons created in a $p+A$ collision have a large probability to decay
into kaon pairs inside the nucleus. These kaons may rescatter
before they leave the nucleus, and consequently the kinematic information
needed to reconstruct the $\phi$ meson might be lost. This effect is
expected 
to be bigger for a larger nucleus, therefore studying the mass dependence
of the number of $\phi$ mesons reconstructed from the $K^+K^-$
channel is a suitable probe for studying the in-medium broadening
of the $\phi$ meson.

$\phi$ meson production in $p+A$ collisions can also be studied via the
dilepton-decay channel $\phi\to e^+ e^-$. Electrons practically do
not interact with the nuclear matter, and the $\phi$ mesons can
be reconstructed using the dilepton invariant mass. The dilepton
decay 
width of the $\phi$ meson is expected to be weakly affected by
the medium. However, the dramatic increase of the dominant kaonic
decay width causes a similar decrease of the dilepton 
branching-ratio, therefore, also the $\phi$ multiplicity observed via the
dilepton channel is affected.

In this paper we present a theoretical study of $\phi$ meson
production in $p+A$ collisions based on calculations carried out using a
BUU type transport model. We study the effect of in-medium $\phi$
meson broadening and kaon rescattering. We give predictions for $\phi$
production cross sections obtained from both the $K^+K^-$ and the
$e^+ e ^-$ channels.

\section{The model}
\label{sec:model}

\subsection{Elementary reactions}
\label{sec:elem}

In addition to $\phi$ meson production in a primary collision
of two nucleons, $NN\to NN\phi$, our model also incorporates the
production of $\phi$ mesons in collisions of secondary particles with
nucleons, namely the channels $\Delta N \to NN\phi$, $\pi N \to
N\phi$ and $\rho N \to N\phi$. These channels have already been
considered for $\phi$ meson production in heavy-ion collisions
in Ref.~\cite{phiAA}, and we utilize the respective  cross sections from
this work.

The total width of the $\phi$ meson is given by
\begin{equation}
  \label{eq:Gamma_tot}
  \Gamma_{tot} = \Gamma_{K^+K^-} + \Gamma_{K^0\bar{K}^0}
  + \Gamma_{e^+e^-} + \Gamma_{\mathrm{rest}},
\end{equation}
where $\Gamma_{\mathrm{rest}}$ means the sum of the widths of the
remaining decay channels. 
The in-medium values of
$\Gamma_{K^+K^-}$ and $\Gamma_{K^0 \bar{K}^0}$ depend 
on the $\phi$ mass and the kaon three-momentum $p_K$ via
\begin{equation}
  \label{eq:Gamma_med}
  \Gamma^\mathrm{med}_{\phi\to K\bar{K}} = \left(
  \frac{m^{\mathrm{vac}}_{\phi}}{m^{\mathrm{med}}_{\phi}}
  \right)^2
  \left(
    \frac{ p^\mathrm{med}_K }
         { p^\mathrm{vac}_K }
  \right)^3 \Gamma^\mathrm{vac}_{\phi\to K\bar{K}}
\end{equation}
for neutral and charged kaon pairs. The superscripts 
`med' and `vac' refer to the in-medium and vacuum values, 
respectively. In vacuum the width $\Gamma_{\mathrm{rest}}$
amounts to 15 \% (0.7 MeV) of the total width and originates
mainly from three-pion decay or the $\pi\rho$ decay. These channels play
a minor role for the change of the total width 
in nuclear matter. E.g., if one assumes that 
50 \% goes through the  $\pi\rho$ channel and the $\rho$ mass is reduced
in matter by 200 MeV the corresponding partial width increases 
to two MeV which is still small compared to the expected change
of the total width. Thus, the variations
of $\Gamma_{\mathrm{rest}}$ would not essentially change our 
results and are therefore neglected in our calculations.
We assume furthermore that $\Gamma_{e^+e^-}$ is not affected by
the medium.

We remark that Eq.~(\ref{eq:Gamma_med}) is a simplification compared to
a rigorous treatment of  the $\phi$ meson spectrum in matter.
This relation would be justified if the spectral function of kaons 
could be approximated with a $\delta$ function. This is justified for 
$K^+$ but hardly for the $K^-$ mesons 
\cite{Kolomeitsev,SibCas,Schaffner,Cieply,Lutz,Kaiser,Oset,Ramos,klingl}.
However, for an attractive $K^-$ potential of 150 MeV 
Eq.~(\ref{eq:Gamma_med}) provides a width of 45 MeV 
which is quite similar to the value obtained 
in Ref.~\cite{klingl}.

The rescattering of the kaons proceeds either via elastic
($K^{\pm}N\to K^{\pm}N$) or inelastic ($K^+ N\to$ $K^+ N\pi$,
$K^- N \to \pi\Lambda$ or $K^- N \to \pi\Sigma$) channels. In our
calculations we used a fit to the available experimental data of the
cross sections of these processes \cite{Landolt-Bornstein}.
We also included the rescattering of the $\phi$ mesons 
and their absorption via the reaction
$\phi N \to$ $K\Lambda$.

\subsection{In-medium potentials}
\label{sec:pot}

The kaonic atom experiments \cite{Katom} indicated that the $K^-$ mass
might decrease by about 200 MeV even at normal nuclear matter
density. Theoretical models that are able to reconstruct the antikaon
enhancement observed 
in $Ni+Ni$ reactions favor a density dependent
potential that gives a smaller mass drop of about 70-120 MeV at
normal nuclear matter density \cite{fopi2,kaos3}. 
The $K^+$ potential has been found to be weakly repulsive and we use
a linear dependence on the baryonic number density $n$
\begin{equation}
  \label{eq:K+pot}
  U_{K^+}(n) = 25 \, \mathrm{MeV} \frac{n}{n_0},
\end{equation}
where $n_0$ is the normal nuclear matter density.
At present there is no experimental information about the $\phi$ meson
potential in the nuclear medium. It is commonly assumed that the
mass weakly depends on the nuclear density.
\begin{equation}
  \label{eq:phipot}
  m_{\phi}^\mathrm{med} = m_{\phi}^\mathrm{vac} \left( 1 - \alpha
  \frac{n}{n_0} \right).
\end{equation}
The parameter $\alpha$  depends  crucially
on the strangeness content of the nucleons. 
Hatsuda et al.\ \cite{Hatsuda} estimated $\alpha$ = 0.025 
while  in refs.~\cite{klingl,cabrera} a 
shift of the $\phi$ mass of about 10 MeV 
and an increase of the width up to 30 MeV was predicted.
The authors of refs.~\cite{kaempfer,zschocke} 
showed using the sum rule approach 
that the mass shift is connected with the
value of the four-quark condensate in the QCD vacuum. 
Reasonable estimates give values up to \mbox{$\alpha$ = 0.033}.
To cover the possible 
 range of the  parameter $\alpha$  we compare  calculations
using  \mbox{$\alpha$ = 0} and  \mbox{$\alpha$ = 0.033}, respectively.

Now we are going to investigate the effects of different in-medium
potentials for the $K^-$ meson.
We carried out four sets of calculations: (a) without an in-medium $K^-$
potential, (b) with a moderate potential derived in Ref.~\cite{Cieply},
(c) with the momentum dependent $K^-$ potential of 
Ref.\ \cite{SibCas}, and (d) with a strong, momentum independent $K^-$
potential. These potentials can be summarized in the form \cite{SibCas}
\begin{equation}
  \label{eq:Kminpot}
  U_{K^-}(n,p_{K}) = 
  \left[a+b \exp(-cp_{K})\right]\frac{n}{n_0}.
\end{equation}
The parameters $a$, $b$ and $c$ of Eq.\ (\ref{eq:Kminpot})
corresponding to these four cases are
\begin{center}
\begin{tabular}{clll}
  (a) & $a$=0, & $b$=0, & $c$=0; \\
  (b) & $a$=$-$70 MeV, & $b$=0, & $c$=0; \\
  (c) & $a$=$-$55 MeV, & $b$=$-$130 MeV, &
     $c$=0.0025 MeV$^{-1}$; \\
  (d) & $a$=$-$150 MeV, & $b$=0, & $c$=0.
\end{tabular}  
\end{center}
The potential (d) corresponds to early estimates to fit the energy levels
in $K^-$ atoms, but it was shown recently \cite{Cieply} that the moderate
potential (b) also satisfies this constraint.

\subsection{The BUU model}
\label{sec:BUU}

The starting point of our calculations is the BUU code used in
\cite{phiAA}. The model has been extended by including the decay of the
$\phi$ meson into a charged kaon pair.  We have
implemented the propagation of the kaon pair which rescatters 
in the nuclear medium  as described in section \ref{sec:elem}. 

To increase statistics we use the perturbative method for $\phi$
production, i.e.\ if in a two-particle collision the threshold is
overcome, the $\phi$ meson is created and weighted with its production
probability. In the case of the $\phi\to$ $K^+K^-$ decay, however, we
made a Monte-Carlo decision in order to avoid producing too large a number
of kaons with very small weights.

We perform calculations for four different systems, $p+C$, $p+Cu$, 
$p+Te$ and $p+Au$, in order to study the dependence on the 
target mass $A$. 
The $\phi$ production cross section was obtained 
integrating over the impact parameter.
We choose a beam energy of
2.5 GeV, which is below the $\phi$ production threshold in a
free nucleon-nucleon collision.
Fermi motion of the nucleons and many-step processes in the target nucleus
will contribute to gain sufficient energy needed for $\phi$ production.

In the calculation we keep
track of the scattering and absorption processes of the $\phi$ mesons 
and the history of the 
kaon pairs if the $\phi$ mesons decay.
The invariant mass of the $K^+K^-$ pairs can deviate from the 
vacuum $\phi$ mass after the kaons
have left the nuclear matter. 
In a static potential the kaons would transform their potential energy
into kinetic energy if they could escape the nucleus potential, and their
invariant mass would take the value of the originally $\phi$ mass. 
A $K^-$ meson resulting from the decay of a very
slow and light $\phi$ meson could not leave the
attractive potential and will finally get absorbed. However, in reality 
the excited target nucleus will expand, therefore the potentials 
depend on time and weakens during the emission process. 
Therefore, the loss (or gain) of the kinetic energy of a kaon does
not generally correspond to the potential at the creation point. Thus, 
the invariant mass of the emitted kaon pairs  are widely spread.
Since the strong attractive $K^-$ potential has the dominant influence,
the invariant 
mass will in most cases exceed the vacuum $\phi$ mass.
We apply the criterion
\begin{equation}
  \label{eq:selection}
  \left| m_{K^+ K^-} - m^{\mathrm{vac}}_{\phi} \right| < 0.05 \,
  \mathrm{ GeV} 
\end{equation}
to select only those kaon pairs from which the $\phi$ meson can be
reconstructed. Thus the $\phi$ multiplicity which is 
experimentally observable via the $K^+K^-$ channel is obtained as 
\begin{equation}
  \label{eq:phimul}
  N_{\phi}^\mathrm{obs} = N_{\phi}^\mathrm{surv} +
  N_{{K}^+{K}^-}^\mathrm{surv}/B_{\phi\to K^+K^-},
\end{equation}
where $N_{\phi}^\mathrm{surv}$ is the number of the $\phi$ mesons per
event that survived,  
$N_{K^+K^-}^\mathrm{surv}$ is the number of kaon pairs 
which fulfill the selection criterion (\ref{eq:selection}) 
and $B_{\phi\to K^+ K^-}$ is the vacuum value of the 
branching ratio of the $\phi$ decay. 
The fact that the cross section depends on 
the selection criterion Eq.~(\ref{eq:selection}) 
due to the spread of the invariant $K^+ K^-$ mass 
was also discussed in Ref.~\cite{muhlich} for photoproduction
of $\phi$ mesons.

We also consider the decay of the $\phi$ mesons into electron pairs. 
The invariant masses of the electron pairs are recorded, and 
Eqs. (\ref{eq:selection},\ref{eq:phimul}) are applied accordingly.
We mention that the long range Coulomb force leads to an
additional spread around the vacuum 
$\phi$ mass  of less than 10 MeV for a heavily charged target 
like {Au}, a value which is
small compared with the expected nuclear effects.

\section{Results}
\label{sec:res}

\begin{figure}
\begin{center}
\includegraphics[width=90mm]{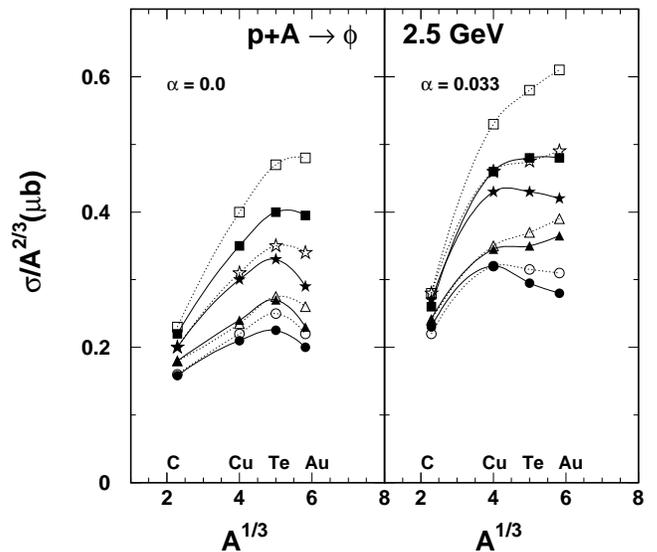}
\end{center}
\caption{
Rescaled cross sections for $\phi$ meson production as
a function of the target mass for different kaon potentials. 
The left hand side picture shows 
the result without changing the $\phi$ mass while on the right hand 
side  the mass diminishes according to Eq.~(\ref{eq:phipot}) with
$\alpha=0.033$. The cross sections are 
 reconstructed from electron pairs (open symbols) and
from kaon pairs (full symbols), respectively.
Squares: without $K^-$ potential (Eq. 5a), 
stars: with a moderate $K^-$ potential (Eq. 5b),
triangles: with momentum dependent potential (Eq. 5c), 
circles: strong momentum independent potential (Eq. 5d).
The lines are included to guide the eyes.
}
\label{sig_a03}
\end{figure}

The main results of our calculations are shown on Fig.\
\ref{sig_a03}. There we display the $\phi$ meson production cross
sections reconstructed from the $K^+K^-$ and the $e^+ e^-$ decay
channels for the four sets of in-medium potentials. The cross
sections are rescaled by $A^{2/3}$ to remove the effect of 
the geometrical cross section of the nucleus.

The comparison of the left and the right hand part of Fig.~\ref{sig_a03}
shows the effect of the change of the  mass of the $\phi$ meson.
The main effect is that the cross section increases if the mass
is diminished in nuclear matter.
That is seen especially
for larger nuclei while the size of a nucleus like 
carbon is not large enough to allow sufficient secondary collision 
inside dense matter.

Without a $K^-$ potential the $\phi$ production cross
section increases with {A} faster than the geometrical cross
section. For target masses below the copper mass the cross
section is roughly proportional to the mass number.
This is because the $\phi$ mesons are predominantly created
in two-step processes by secondary $\pi$ and $\rho$ mesons. 
For larger nuclei the increase is moderate because $\phi$ mesons
get absorbed in large nuclei. This is obvious if the decay
is hindered inside the nucleus because of a reduction of the
mass (see right hand side of  Fig.~\ref{sig_a03}).
We observe an enhancement of the
dilepton channel relative to the $K^+K^-$ channel. This enhancement
is largest if the $K^-$ potential vanishes and is
explained by the fact that the dilepton branching-ratio 
increases with dropping $\phi$ meson mass and  increasing
$K^+$ mass.

A strong attractive $K^-$ potential causes a
rapid decay of the $\phi$ mesons inside the nucleus. This effect 
increases with the size of the target nucleus and is seen in both
the numbers of observed $K^+K^-$ and the $e^+ e^-$ pairs. 
In the case of the $K^+K^-$ channel the reduction of the $\phi$
cross section is caused by the rescattering of the kaons
while in the case of the $e^+ e^-$ channel a similar reduction
occurs because of the decrease of the dilepton branching-ratio
in nuclear matter.

\section{Conclusions}

The results show that the target mass dependence of the observed $\phi$
production cross section is strongly altered if strong 
changes of the kaon masses in nuclear matter occur. 
If such effects exist they  should be measurable 
by detecting kaon or electron pairs in subthreshold
proton-nucleus reactions with energies available at the COSY or SIS 
accelerators at FZ J\"ulich and at GSI Darmstadt, respectively.

\section{Acknowledgement}
M.Z. acknowledges the warm hospitality of the theory group
of the FZ Rossendorf and the financial support of the
S\"achsische Staatsministerium f\"ur Wissenschaft und Kunst.
The work is supported by BMBF 06DR121 and the Bergen Computational Physics
Laboratory.

\end{document}